\documentclass[]{spie}  

\usepackage{amsmath,amsfonts,amssymb}
\usepackage{textcomp}
\usepackage{graphicx}
\usepackage[colorlinks=true, allcolors=blue]{hyperref}

\title{Development Status of Wideband Millimeter-Wave Receivers for LMT-FINER}

\author[a]{Haoran Kang}
\author[a]{Takafumi Kojima}
\author[b]{Takeshi Sakai}
\author[c]{Yoichi Tamura}
\author[b]{Airi Tetsuka}
\author[a]{Sho Masui}
\author[d]{Tatsuya Takekoshi}

\affil[a]{National Astronomical Observatory of Japan, Mitaka, Tokyo 181-8588, Japan}
\affil[b]{University of Electro-Communications, Chofu, Tokyo 182-8585, Japan}
\affil[c]{Nagoya University, Furo, Chikusa, Nagoya 464-8602, Japan}
\affil[d]{Kitami Institute of Technology, Kitami, Hokkaido 090-8507, Japan}

\authorinfo{Further author information: (Send correspondence to H.K)\\H.K: E-mail: haoran.kang@nao.ac.jp\\}

\begin{document} 
\maketitle

\begin{abstract}
Spectroscopic observations of the far-infrared [\textsc{O\,iii}] 88~$\mu$m and [\textsc{C\,ii}] 158~$\mu$m lines present a pathway to explore the mechanisms of the emergence of massive galaxies in the epoch of reionization and beyond, which is one of the most fundamental questions in astronomy. To address this question, the Far-Infrared Nebular Emission Receiver (FINER) project is developing two wideband dual-polarization sideband-separating heterodyne receivers at 120--210 GHz and 210--360 GHz for the Large Millimeter Telescope (LMT) in Mexico. 
Compared with Atacama Large Millimeter/submillimeter Array (ALMA), LMT provides 40\% of ALMA's light-collecting area and a similar atmospheric transmittance, but FINER plans to have an instantaneous intermediate frequency (IF) of 3--21 GHz per sideband per polarization which is five times wider than current ALMA's bandwidth. Therefore, FINER is going to offer cutting-edge spectral scanning capability in the next several years.

The project is currently in an active development phase. In this proceeding, the latest development status for FINER, including the optics, wideband waveguide components as well as low-noise superconductor-insulator-superconductor (SIS) mixers is reported.
\end{abstract}


\section{INTRODUCTION}
\label{sec:intro}  
When massive galaxies emerge and how common they are in the epoch of reionization and beyond is one of the most fundamental questions in modern astronomy. Although recent observations by James Webb Space Telescopes (JWST) suggest an exceptionally efficient galaxy assembly process in the earliest universe, the physical mechanisms driving their rapid growth remain an unsolved puzzle. 
Recent studies show that redshifted far-infrared [\textsc{O\,iii}] 88~$\mu$m line is a very useful tracer to identify the spectroscopic redshifts of the earliest massive galaxies in the epoch of reionization and beyond~\cite{inoue2016detection,hashimoto2018onset,tamura2019detection}, and together with the observations of [\textsc{C\,ii}] 158~$\mu$m line, it is possible to  further elucidate their physical properties, such as gas-phase metallicity, ionizing photon and gas densities, and multi-phase gas structure, which are the key to unveil the origins of the extraordinary efficiency of star formation in these galaxies~\cite{hashimoto2019big,bakx2020alma,harikane2020large}.
Therefore, the (sub)millimeter-wave spectroscopic observations of [\textsc{O\,iii}] 88~$\mu$m and [\textsc{C\,ii}] 158~$\mu$m lines provide a pathway to unravel the fundamental question of first galaxies.

To cover both redshifted [\textsc{O\,iii}] and [\textsc{C\,ii}] lines in a large redshift range, low-noise heterodyne receivers with ultra-wide radio frequency (RF) bandwidths are required. In addition, because of the large uncertainty of the photometric redshift of galaxy candidates, wide intermediate frequency (IF) bandwidths are also critical for these observations. However, currently, (sub)millimeter-wave instruments with required sensitivity and bandwidths are very limited in the world. 
To address this challenge, the Far-Infrared Nebular Emission Receiver (FINER) project~\cite{tamura2024finer} proposed to develop low-noise dual-polarization sideband-separating heterodyne receivers that cover the frequency range of 120--350 GHz for the Large Millimeter Telescope (LMT)~\cite{hughes2010large,hughes2020large} in Mexico with a best effort to extend the frequency range to 360 GHz.
Thanks to the recent developments in the ultra-wideband waveguide components~\cite{gonzalez2017recent,gonzalez2018double,yagoubov2020wideband,gonzalez2021high,masui2021development,kang2023wideband} and SIS mixers\cite{kojima2017performance,kojima2018275,kojima2020wideband}, only two heterodyne receivers are sufficient to cover such a wide frequency range. In specific, one receiver covers 120--210 GHz which is corresponding to the ALMA Band 4+5 frequencies, another one covers 210--360 GHz which is corresponding to the ALMA Band 6+7 frequencies~\cite{wootten2003atacama}. 
In this proceeding, the frequency band definition of ALMA is also applied.
Because both receivers have fractional bandwidths larger than 50\%, a great deal of effort has been devoted for the development of the wideband waveguide components and SIS mixers. 
In comparison to ALMA, these receivers plan to provide an IF bandwidth of 3--21 GHz per sideband per polarization, providing 40\% of ALMA's light-collecting area and a similar atmospheric transmittance, but a five times wider bandwidth than current ALMA's bandwidth~\cite{wootten2003atacama}. In the coming years, FINER will offer cutting-edge spectral scanning capability for the study of early universe.

The achievement of our scientific goals depends on an excellent sensitivity of the instrument, demanding both high aperture efficiency and low system noise temperature. Receiver optics determines the aperture efficiency which also affects the system noise temperature. On the other hand, waveguide components include corrugated horns, ortho-mode transducers (OMT) and 2SB units play an important role in minimizing reflection, polarization leakage and improving image rejection ratio across a wide frequency range. In this proceeding, the development status of these components will be reported. Furthermore, the development status of superconductor-insulator-superconductor (SIS) mixers will be also presented, which are the most critical devices for high sensitivity (sub)millimeter-wave heterodyne receivers.

\section{Optics design} 

\begin{figure} [ht]
   \begin{center}
   \begin{tabular}{c} 
   \includegraphics[height=8cm]{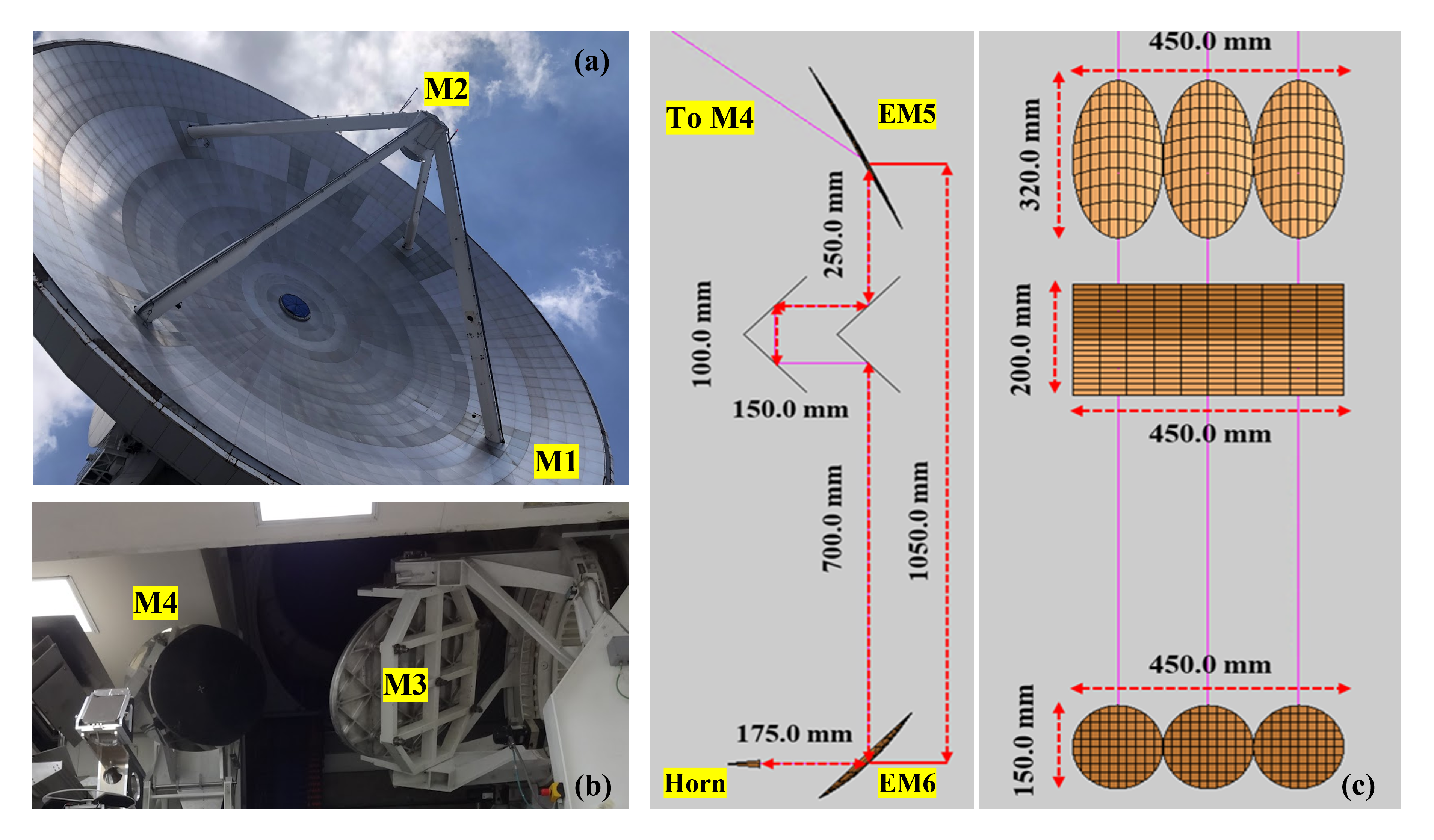}
   \end{tabular}
   \end{center}
   \caption[FINER optics] 
   { \label{fig:FINER optics} 
(a) The primary reflector (M1) and secondary reflector (M2) of LMT. (b) The third and fourth flat mirrors (M3, M4)of LMT. (c) The FINER optics. Additional set of ellipsoidal mirrors is prepared for the future upgrade of 3--4 mm receiver~\cite{tamura2024finer}.}
   \end{figure}

LMT is the largest single-dish millimeter-wave telescope which is capable of efficient astronomical observations up to 1mm wavelength~\cite{hughes2020large}. 
It is a Cassegrain antenna system with a 50m-diameter primary reflector. Besides the secondary reflector, LMT also has two large pick-up mirrors (M3 and M4) to guide the beams to the receiver cabin which has a huge space for different scientific instruments (see Fig.~\ref{fig:FINER optics}). The basic optical parameters of LMT are summarized in Table.~\ref{tab:LMT optics}. 

\begin{table}[ht]
\caption{Summary of the basic optical parameters of the Large Millimeter Telescope and FINER receiver optics. The $f/D$ ratio of the primary reflector is 0.35, the magnification of LMT is 30, and the equivalent focal length of LMT is 525 m ~\cite{olmi1998large}. $f$ is the focal length, $d$ is the distance between each mirror.} 
\label{tab:LMT optics}
\begin{center}       
\begin{tabular}{ccccc} 
\hline
\rule[-1ex]{0pt}{3.5ex}  Surface & Type & Aperture (mm) & $f$ (mm) & $d$ (mm) \\
\hline
\rule[-1ex]{0pt}{3.5ex} M1 & Paraboloidal & 50000.0 & 17500.0 & 0.0 \\
\rule[-1ex]{0pt}{3.5ex} M2 & Hyperboloidal & 2630.0 & -884.3 & 16645.2 \\
\rule[-1ex]{0pt}{3.5ex} M3 & Flat & 1670.0 $\times$ 1100.0 &  & 22895.2  \\
\rule[-1ex]{0pt}{3.5ex} M4 & Flat & 1252.0 $\times$ 880.0 &  & 1500.0\\
\rule[-1ex]{0pt}{3.5ex} EM5 & Ellipsoidal & 160.0 $\times$ 150.0 & 475.2 & 2200.0 \\
\rule[-1ex]{0pt}{3.5ex} EM6 & Ellipsoidal & 150.0 $\times$ 150.0 & 145.6 & 1350.0 \\
\hline
\end{tabular}
\end{center}
\end{table}

The basic function of FINER optics is to illuminate the primary reflector with a 12 dB edge taper which is an optimum value to realize an aperture efficiency above 80\%. As shown in Fig.~\ref{fig:FINER optics}(c), two ellipsoidal mirrors for Band 4+5 frequencies and Band 6+7 frequencies are designed based on the quasi-optical theory~\cite{goldsmith1998quasioptical} to meet the frequency-independent condition~\cite{chu1983imaging, gonzalez2016frequency}. The basic optical parameters of these two ellipsoidal mirrors are also presented in Table.~\ref{tab:LMT optics}. Because of the presence of the fast sky chopper which is describe in the following context, the optics design of the receivers is identical. 
Furthermore, the Band 4+5 and Band 6+7 corrugated horns also have a same aperture diameter of 11.9 mm, and a slant length of 53.9 mm.
In principle, by complying with the frequency-independent condition, it is possible to realize a constant illumination on the primary reflector in a wide frequency range. In this case, if the truncation losses of other mirrors are negligible, the aperture efficiency can be almost constant. 
A preliminary physical optics simulation shows that the aperture efficiency at 210-360 GHz frequencies is better than 80\%, however it degrades to 74\% at 120 GHz which is the low frequency limit of the Band 4+5 receiver. It is mainly due to the relatively large truncation at low frequencies. Because of the limited available space of FINER's cryostat, it is hard to further improve the aperture efficiency by increasing the size of the mirrors. Although the aperture efficiency is lower than 80\% at low frequencies of Band 4+5 receiver, it still represent a good performance.

However, aperture efficiency is not the only consideration for FINER project. The sky subtraction at Band 6+7 frequencies is very challenging due to the unstable atmospheric noise. Therefore, a beam switching mechanism called the fast sky chopper will be applied in FINER receivers. The fast sky chopper is a set of removable mirrors to switch the pointing direction on the sky of each beam at a frequency around 1--2 Hz. This device allows beam-switching observations which are more efficient for sky subtraction compared to conventional position-switching observations~\cite{tamura2024finer}. Considering the complexity of the fast sky chopper design, it is expected that this device will not be installed during the first-light observations of FINER. It will be implemented as a future upgrade. 

Because the fast sky chopper changes the optical length between the two ellipsoidal mirrors, to keep the optical parameters of these mirrors in constant before and after the installation of the fast sky chopper, four additional flat mirrors need to be added between the mirrors, which change the optical length by 300.0 mm. Large number of mirrors are not preferred in low-noise heterodyne receivers because of the contribution of extra noise temperature~\cite{lamb2003low}. Based on the method in Ref.~\citenum{walker2015terahertz}, the extra noise temperature of the four additional mirrors is analyzed, it shows that at 120--210 GHz, this noise contribution is around 5 K, and at 210--360 GHz, it increase from $\sim$ 5 K to $\sim$ 9 K at the high frequencies due to the worse atmospheric transmission. 
Although it is not negligible, the extra noise from the additional four flat mirrors is pretty small in comparison to other noise sources.

\section{Wideband waveguide components}

\subsection{Waveguide components at 210--360 GHz}
   
The development of the key wideband waveguide components at Band 6+7 frequencies has been completed by the end of May 2024, including high-performance corrugated horn, ortho-mode transducer (OMT) and two sideband (2SB) units. In this subsection, the development status of these components including the comprehensive measurement results is described.
\subsubsection{Corrugated horn}

   

A corrugated horn is an important component in the waveguide circuit to couple the radiation from the optical system to the waveguide circuit with high efficiency for an astronomical receiver. 
The aperture diameter and slant length of the Band 6+7 corrugated horn is decided by the quasi-optical design of FINER optics. Therefore, the goal of the electromagnetic design of the corrugated horn is to optimize the reflection loss and the cross-polarization performance.

   \begin{figure} [ht]
   \begin{center}
   \begin{tabular}{c} 
   \includegraphics[height=13cm]{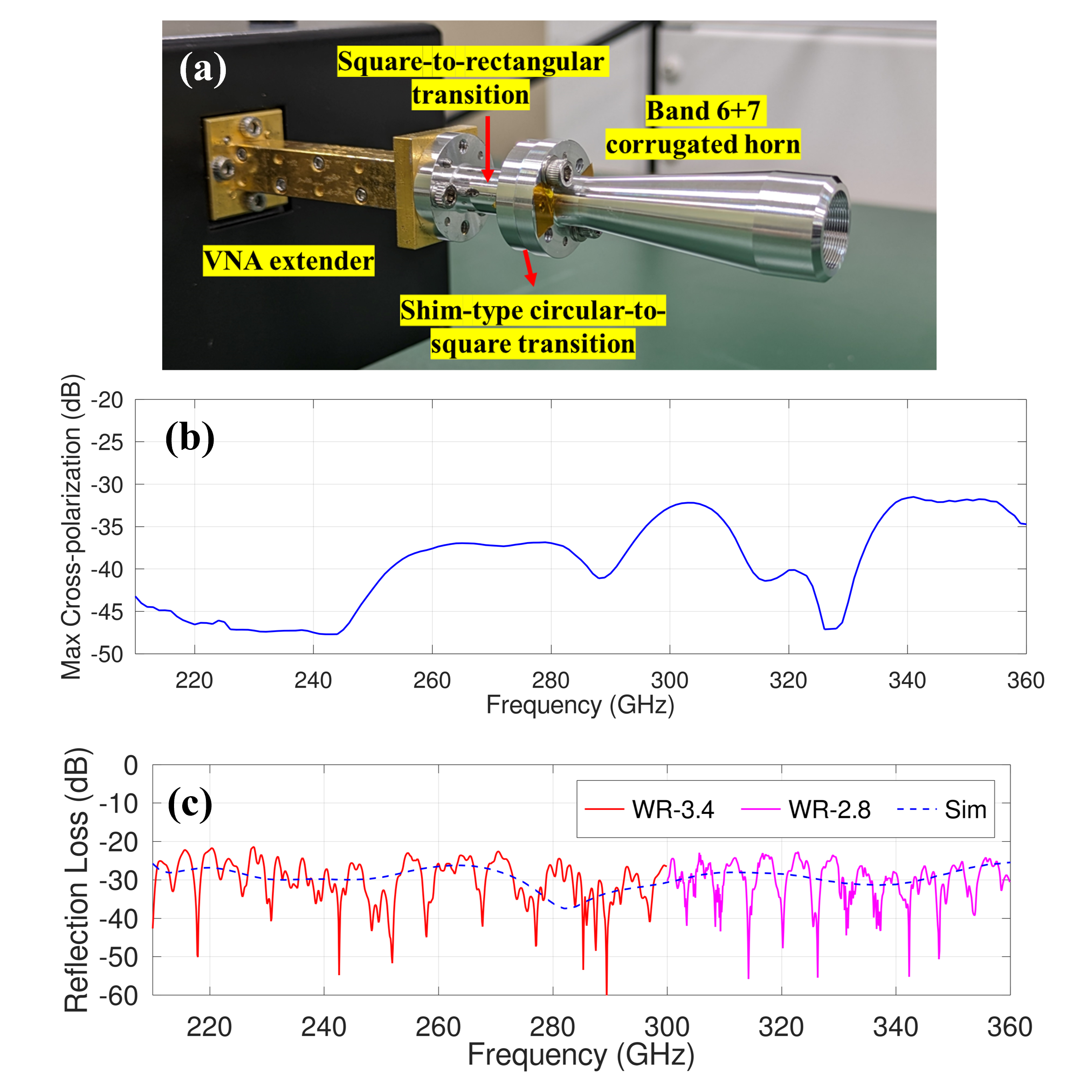}
   \end{tabular}
   \end{center}
   \caption[B6+7 horn + meas] 
   { \label{fig:B6+7 horn + meas} 
(a) The measurement setup for the Band 6+7 corrugated horn. (b) Maximum cross-polarization in the simulated far-field beam patterns of the Band 6+7 corrugated horn. (c) The measured reflection loss of the Band 6+7 corrugated horn.}
   \end{figure}
   
In terms of reflection loss, previous studies, such as Ref.~\citenum{zhang1993design} and Ref.~\citenum{gonzalez2017recent}, have presented the basic methodology for the optimization of the reflection performance for ultra-wideband corrugated horns. 
In the design of the Band 6+7 corrugated horn, the slot depths of first ten corrugations and the slot widths of first six corrugations have been fully optimized by a commercial hybrid mode-matching (MM)/finite element (FE) software Wasp-NET~\cite{arndt2012wasp} to achieve a reflection loss lower than -23.9 dB in the whole frequency range.
On the other hand, low-level of cross-polarization (XsP) is another important aspect for high-performance corrugated horns. As explained in Ref.~\citenum{clarricoats1984corrugated}, the slot depth for the minimum XsP of hybrid HE$_{11}$ mode is a function of the radius of corrugated waveguide, and this depth should increase when the corrugated waveguide has a smaller radius.
Because a corrugated horn consists of corrugated waveguides with different radii, profiled slots which have a form of Equation~(\ref{eq:slot}) is proposed in Ref.~\citenum{kang2023wideband}.

\begin{equation}
\label{eq:slot}
s = a_1 \left(\frac{2r}{\lambda}\right)^{-a_2}+a_3 ,
\end{equation}
where $r$ is the radius of corrugated waveguide, $\lambda$ is the wavelength, and $a_1$, $a_2$, $a_3$ are the free parameters that should be optimized. 
In order to achieve the best possible XsP performance for the corrugated horn, the three parameters in Equation~(\ref{eq:slot}) are used to optimize the XsP of the Band 6+7 corrugated horn. As shown in Fig.~\ref{fig:B6+7 horn + meas}(b), finally a XsP-level lower than -31.5 dB has been realized in the whole frequency range. Because the corrugated horn has a circular waveguide input, a shim-type circular to square waveguide transition also has been carefully designed to connect the corrugated horn with the square waveguide input of the OMT\cite{gonzalez2023practical}. This shim-type transition has a reflection loss better -30.0 dB and negligible insertion loss because of its very short length. As shown in Fig.~\ref{fig:B6+7 horn + meas}(a), this transition can be simply inserted as a shim between the corrugated horn and other waveguide components.

After the design, the Band 6+7 corrugated horn has been fabricated by Kawashima Manufacturing in Japan~\cite{kawashima}. Then, the reflection loss of the horn has been characterized by a Keysight PNA-X Vector Network Analyzer (VNA) at Advanced ICT Device Laboratory in National Institute of Information and Communications Technology (NICT). Because the fractional bandwidth of this horn is very wide (52.7\%), VNA extenders at two different frequency ranges are used for the measurement. In specific, VDI WR-3.4 extenders are used for the measurement at 210--300 GHz, and VDI WR-2.8 extenders are used for the measurement at 300--360 GHz.
As shown in Fig.~\ref{fig:B6+7 horn + meas}(c), the reflection loss of this corrugated horn is lower than -21.4 dB in the whole frequency range, which is an excellent performance compared to the state-of-the-art ultra-wideband corrugated horns at similar frequency ranges~\cite{lee2023development,hu2021design}. Although the far-field beam patterns have not been measured by the time of the conference, it is scheduled soon at the anechoic chamber in NICT.

\subsubsection{Ortho-mode transducer}

\begin{figure} [h]
   \begin{center}
   \begin{tabular}{c} 
   \includegraphics[height=5cm]{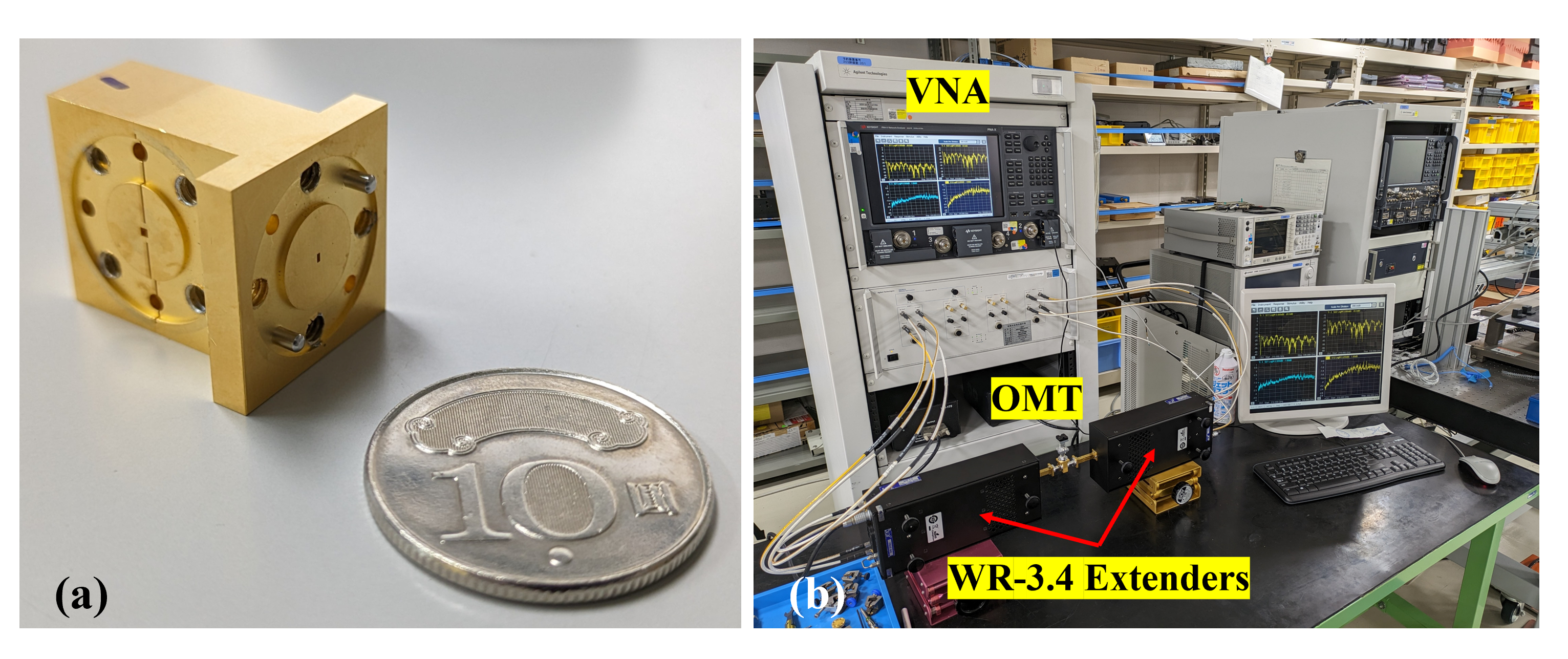}
   \end{tabular}
   \end{center}
   \caption[B6+7 OMT] 
   { \label{fig:B6+7 OMT} 
(a) The Band 6+7 OMT. (b) The measurement setup for the Band 6+7 OMT. It inludes a Keysight PNA-X VNA and a set of VNA extenders (only WR-3.4 extenders are shown in this figure).}
   \end{figure}

The design of the Band 6+7 OMT is provided by Osaka Metropolitan University, and more details of this OMT are described in Ref.~\citenum{masui2024receiver}. Then the OMT is fabricated by Kawashima Manufacturing in Japan~\cite{kawashima}.
Because ohmic loss of metallic materials depends on the equivalent conductivity, which is highly affected by the surface roughness, gold plating is applied on the surface of this OMT to further improve the surface roughness of direct machined aluminum for a lower insertion loss (see Fig.~\ref{fig:B6+7 OMT}(a)).

   \begin{figure} [ht]
   \begin{center}
   \begin{tabular}{c} 
   \includegraphics[height=14cm]{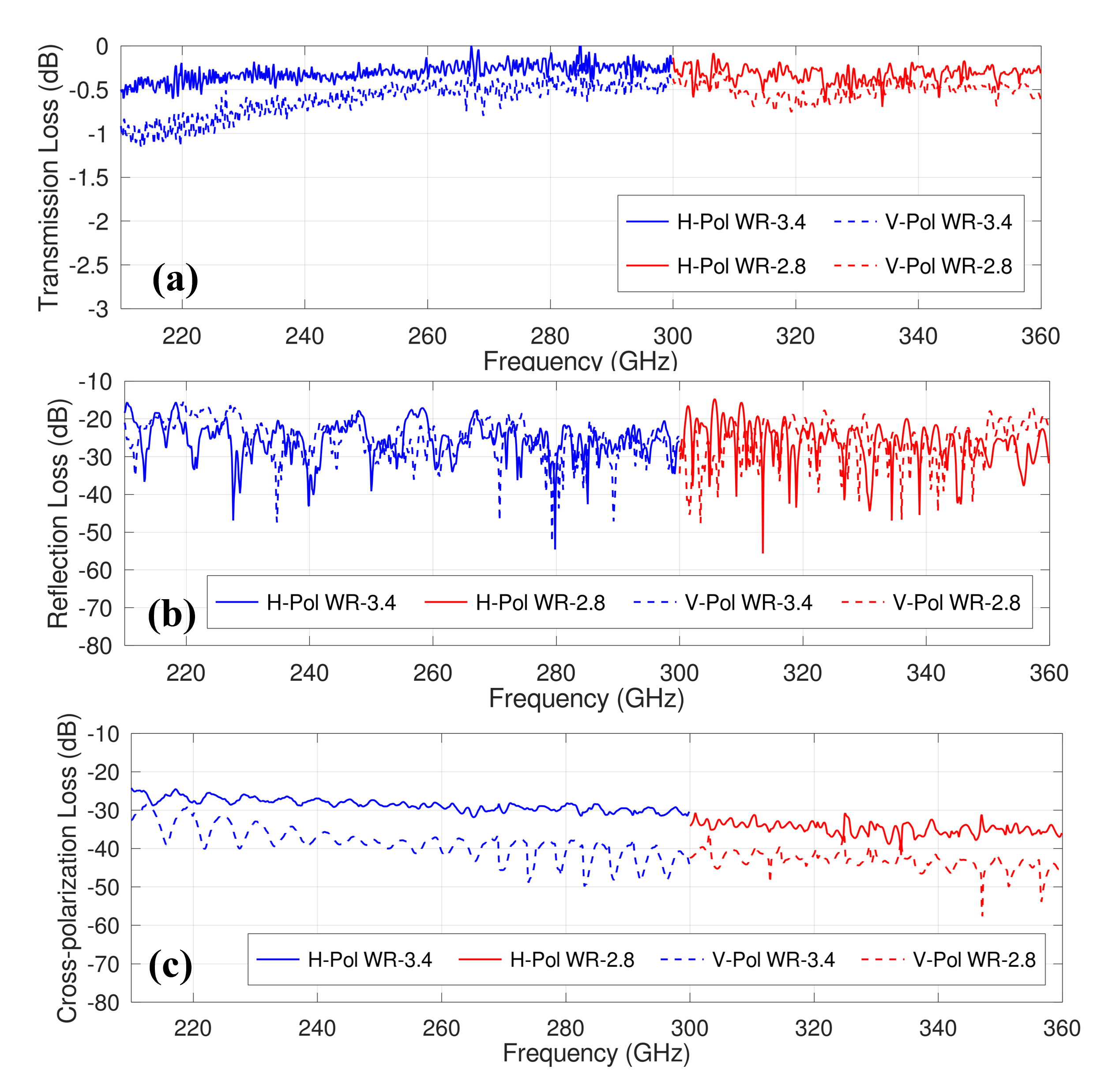}
   \end{tabular}
   \end{center}
   \caption[B6+7 OMT meas] 
   { \label{fig:B6+7 OMT meas} 
(a) The measured transmission loss of the Band 6+7 OMT at room temperature. (b) The measured reflection loss of the Band 6+7 OMT. (c) The measured cross-polarization loss of the Band 6+7 OMT.}
   \end{figure}
As shown in Fig.~\ref{fig:B6+7 OMT}(b), this OMT has also been characterized by a Keysight PNA-X VNA at NICT Advanced ICT Device Laboratory. As mentioned in the previous context, VNA extenders at two different frequency ranges are used for the measurement. To connect the VNA extenders to the OMT with low reflection, two different kinds of rectangular waveguide transitions and square waveguide transitions are also carefully designed and fabricated for this measurement.

The measurement result is presented in Fig.~\ref{fig:B6+7 OMT meas}. This measurement is based on the methodology presented in Ref.~\citenum{navarrini2021}. 
Because the parameter of transmission loss is also used in this proceeding, it is worth mentioning that the sign of the insertion loss is opposite to the transmission loss when they are both expressed in decibel.
The measured transmission loss is around -0.5 -- -0.3 dB for horizontal polarization (H-pol) and -1.0 -- -0.4 dB for vertical polarization (V-pol). The transmission loss for V-pol is lower than H-pol because of the longer waveguide length of V-pol, which is limited by the flange size of square waveguide input. 
The measurement is conducted at room temperature. As presented in Ref.~\citenum{garrett2022measuring}, the equivalent conductivity of gold-plated aluminum will increase when it is cooled down to 4 K, therefore the transmission loss of this OMT can be further improved in cryogenic temperatures.
In terms of reflection loss, the measurement shows a maximum reflection loss of -14.7 dB for H-pol and -15.0 dB for V-pol. The relatively high measured reflection loss is considered partly due to the poor reflection performance of the square waveguide transitions.
Because polarization discrimination capability is the most important function of OMTs, the cross-polarization loss of Band 6+7 OMT has also been carefully measured. The measurement shows a maximum cross-polarization of -24.2 dB for H-pol and -28.5 dB for V-pol, which is a good cross-polarization level at this frequency range. The degradation of cross-polarization performance is mainly due to the misalignment ($\sim$ 5$\mu$m) between the two split blocks of OMT, and it is difficult to make significant improvements under the existing high-precision machining technology.

\subsubsection{2SB unit}
   


\begin{figure} [ht]
   \begin{center}
   \begin{tabular}{c} 
   \includegraphics[height=13.5cm]{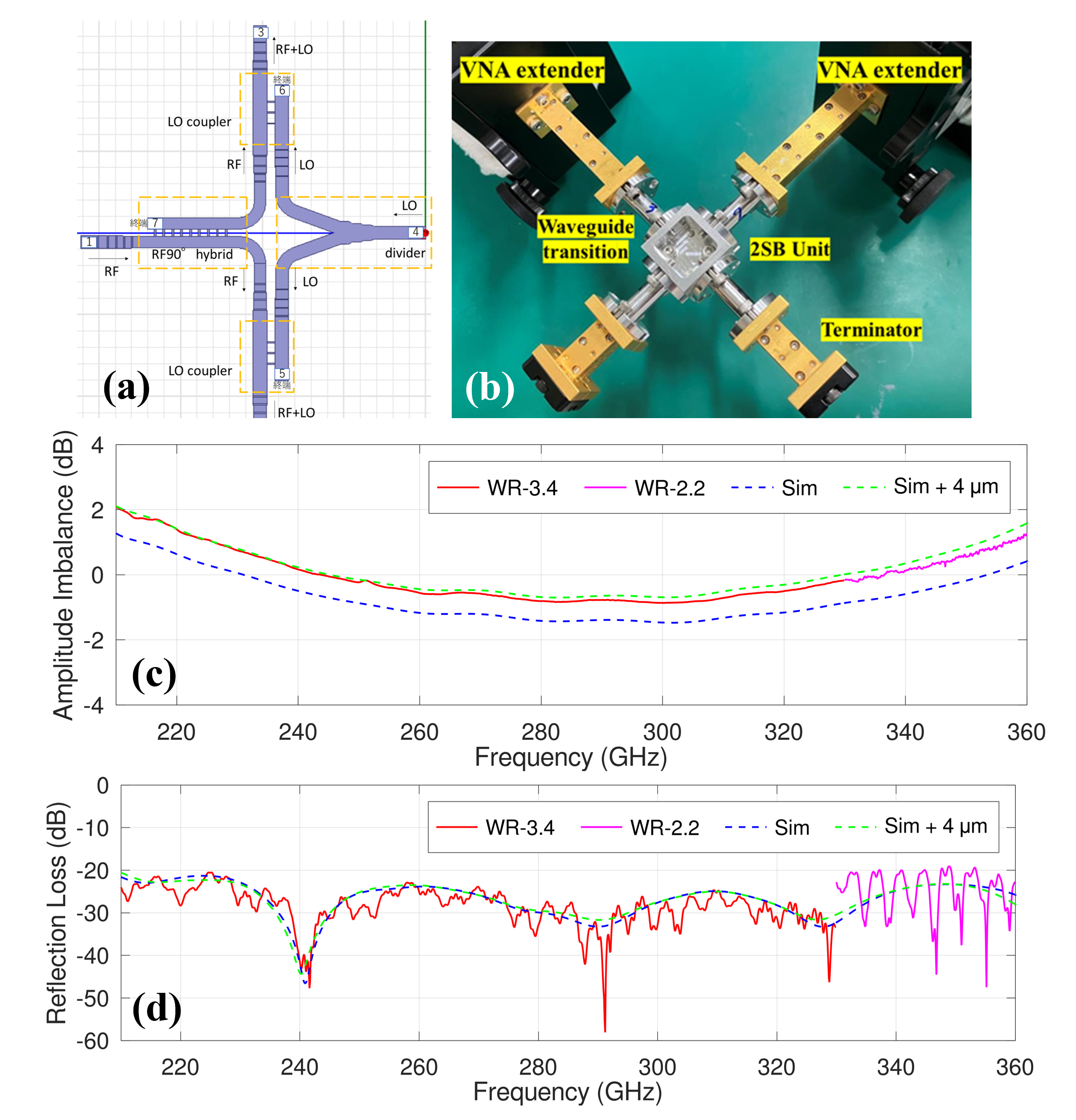}
   \end{tabular}
   \end{center}
   \caption[B6+7 2SB meas] 
   { \label{fig:B6+7 2SB meas} 
(a) The HFSS simulation model of the Band 6+7 2SB unit. (b) The measurement setup for the Band 6+7 2SB unit. (c) The measured amplitude imbalance of the Band 6+7 2SB unit. (d) The measured reflection loss of the Band 6+7 2SB unit.}
   \end{figure}

A 2SB unit comprises of a 90-degree hybrid coupler, a local oscillator (LO) power divider and two LO couplers. It provides the 90-degree phase difference for sideband separation and the necessary LO power to pump the SIS mixers. The amplitude imbalance and phase difference of the 90-degree hybrid coupler significantly affects the image rejection ratio (IRR), which indicates the response to undesirable sideband signals.
Since IRR is more sensitive to the amplitude imbalance, in the design of 90-degree hybrid couplers, narrow slots are always used to improve the amplitude imbalance, however there is always a limitation of the slot width that can be properly fabricated. 
As shown in Fig.~\ref{fig:B6+7 2SB meas}(a), in the design of the Band 6+7 2SB unit, 9 slots with a slot width of 60 $\mu$m are used to achieve a maximum amplitude imbalance lower than 1.5 dB, and the phase difference between the two ports of the 90-degree hybrid coupler is within 90\textdegree $\pm$1\textdegree. 

Two Band 6+7 2SB units that are needed for the Band 6+7 receiver have been fabrication by Kawashima Manufacturing in Japan~\cite{kawashima}. Thanks to the symmetric geometry of the 2SB unit, the component is fabricated by direct machining of aluminum in two split blocks. Then, one of them has been characterized by a Keysight PNA-X VNA at Advanced Technology Center (ATC) in NAOJ. As mentioned previously, because of the wide fractional bandwidth, VDI WR-3.4 extenders are used for the measurement at 210--330 GHz, and VDI WR-2.2 extenders are used for the measurement at 330--360 GHz. As shown in Fig.~\ref{fig:B6+7 2SB meas}(b), the maximum amplitude imbalance is around 2.0 dB, which is slightly worse than the design value. After the measurement of the slot widths of the 90-degree hybrid coupler, it is confirmed that the slot widths are 4 $\mu$m larger than the design value. A further investigation has been done by conducting a HFSS~\cite{hfss} simulation with wider slot widths, and the comparison between simulation and measurement results is also shown in Fig.~\ref{fig:B6+7 2SB meas}(b). It shows that wider slot widths degrade the amplitude imbalance in the fabricated component, and the measurement data closely fits the simulation result.
Although the larger amplitude imbalance potentially affects the IRR performance of the Band 6+7 receiver, digital sideband separation will be applied in the backend spectrometer of FINER receivers, and the amplitude imbalance can be compensated in the digital process \cite{finger2013calibrated,hagimoto2024spectrometer}.

Furthermore, to minimize the reflection of the 2SB unit, the impedance transformers between each element of the 2SB unit are also carefully optimized. 
The measured reflection loss is shown in Fig.~\ref{fig:B6+7 2SB meas}(c), it shows a reflection loss better than -19.0 dB in the whole frequency range. It clearly shows in the figure that the higher reflection appears in the frequency range of WR-2.2 extenders, which indicates the high-level of reflection may be due the extra reflection of WR-2.2 rectangular waveguide transitions. 
Although the reflection loss is slightly worse than the design, it still demonstrates an excellent reflection performance of this component.

\subsection{Waveguide components at 120--210 GHz}
By the time of the conference, the Band 4+5 OMT has been developed and fully characterized, and the measurement result will be presented in the following part of this section. Regarding other wideband waveguide components, the development work of corrugated horn and 2SB unit is still ongoing. 

The Band 4+5 corrugated horn has the same aperture diameter and slant length as the Band 6+7 corrugated horn due to the design of FINER	optics. On the other hand, the slot depths and slot widths of corrugations for this horn are adjusted for the lower frequency range. By following the same optimization strategy of the Band 6+7 corrugated horn, reflection loss lower than -24.7 dB and XsP-level lower than -33.9 dB has been realized. The fabrication of this corrugated horn is scheduled to be in the second half of 2024 by Kawashima Manufacturing~\cite{kawashima}. 
The initial design of the Band 4+5 2SB unit is a scale model of the Band 6+7 2SB unit, the target specifications of amplitude imbalance and phase difference are same as the ones for 210--360 GHz, which are 1.5 dB and 90\textdegree $\pm$1\textdegree, respectively. 
Currently, the dimensions of the Band 4+5 2SB unit are optimized by Wasp-NET~\cite{arndt2012wasp} for a better reflection performance. The design of this component will be completed in the next several months, and the fabrication will be in the second half of 2024.

\subsubsection{Ortho-mode transducer}

   \begin{figure} [ht]
   \begin{center}
   \begin{tabular}{c} 
   \includegraphics[height=10cm]{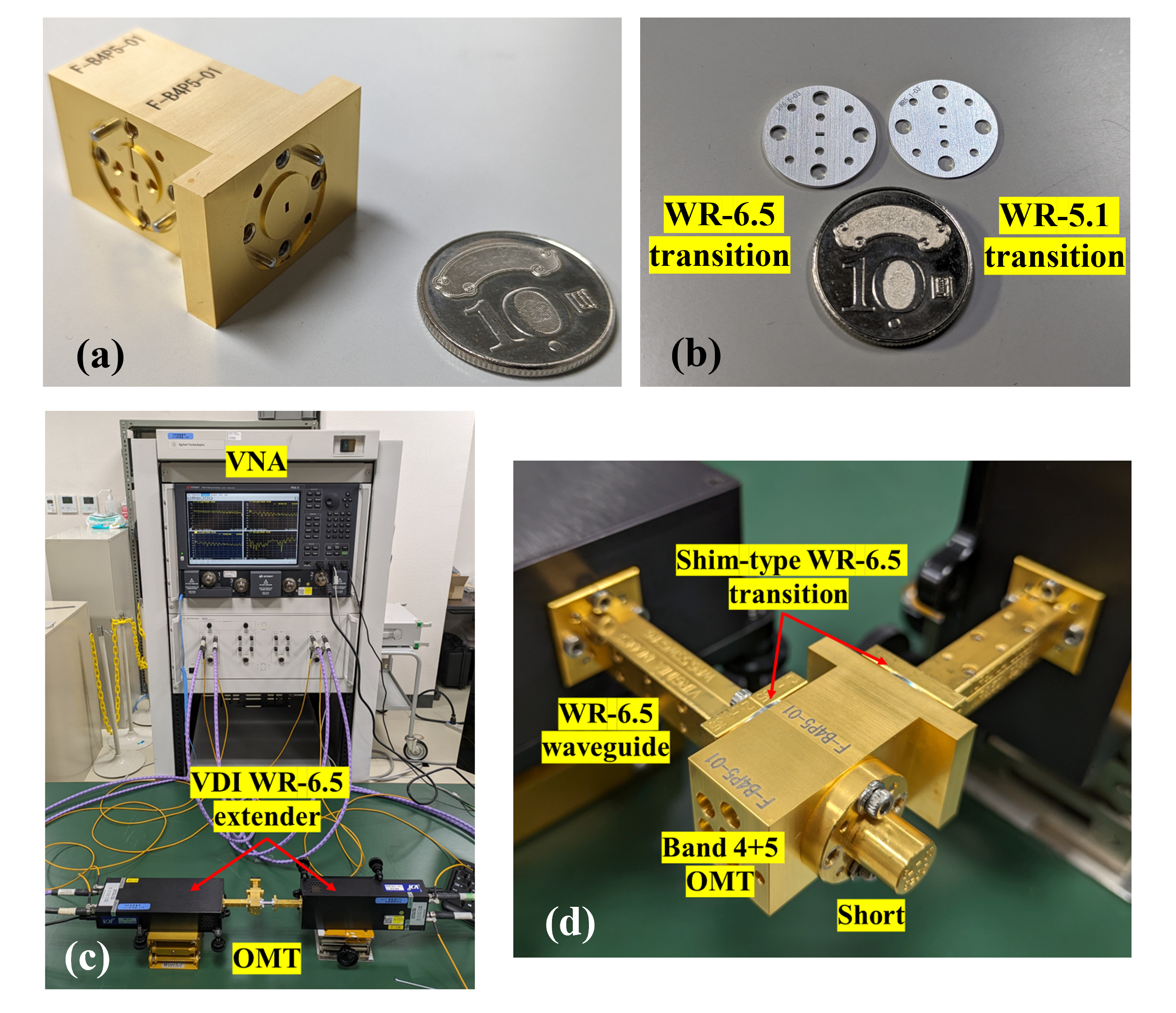}
   \end{tabular}
   \end{center}
   \caption[B4+5 OMT] 
   { \label{fig:B4+5 OMT} 
(a) The Band 4+5 OMT. (b) Two kinds of shim-type rectangular waveguide transitions to cover 120--210 GHz frequency range. (c) The measurement setup for Band 4+5 OMT. It inludes a Keysight PNA-X VNA and a set of extenders (only WR-6.5 extenders are shown in this figure). (d) A close view of the connection between the OMT and the VNA extenders.}
   \end{figure}

   \begin{figure} [h]
   \begin{center}
   \begin{tabular}{c} 
   \includegraphics[height=14cm]{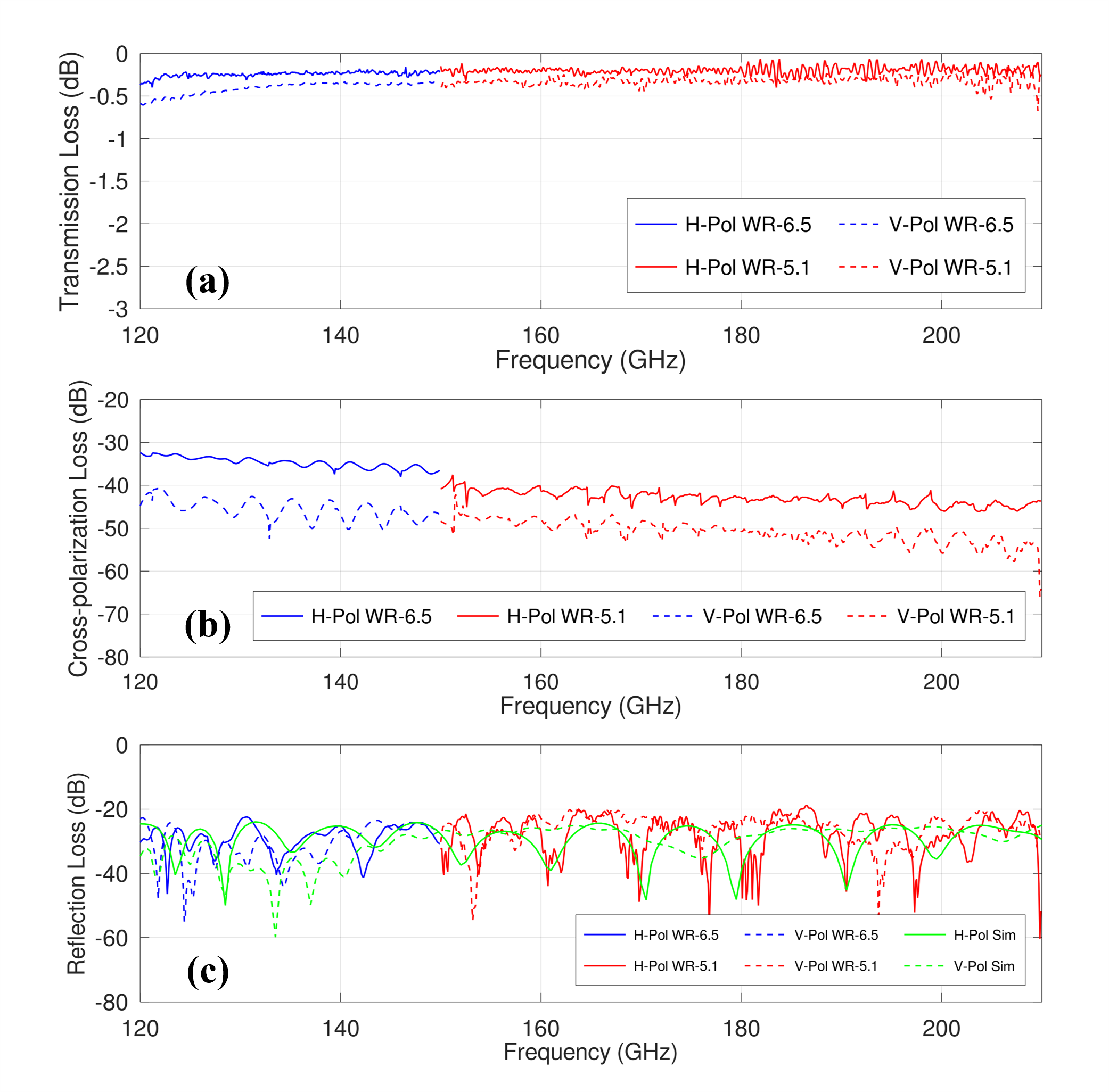}
   \end{tabular}
   \end{center}
   \caption[B4+5 OMT Meas] 
   { \label{fig:B4+5 OMT Meas} 
(a) The measured transmission loss of the Band 4+5 OMT at room temperature. (b) The measured cross-polarization loss of the Band 4+5 OMT. (c) The measured reflection loss of the Band 4+5 OMT.}
   \end{figure}
   
   
A double-ridged waveguide OMT design has been chosen for Band 4+5 frequencies because of their good insertion loss performance compared to turnstile junction OMTs and low intrinsically cross-polarization loss. In recent years, this kind of OMTs have been widely used for ultra-wideband (sub)millimeter-wave heterodyne receivers. Detailed description of how to design such kind of high-performance OMTs can be found in Ref.~\citenum{gonzalez2018double}, Ref.~\citenum{gonzalez2021high} and Ref.~\citenum{gonzalez2023practical}. 

At first, the dimension of the H-pol and V-pol rectangular waveguide outputs of the OMT is chosen to be 1.410 $\times$ 0.705 mm (WR-5.6).
Therefore, the cutoff frequency of fundamental TE$_{10}$ mode is 106.3 GHz, and the cutoff frequency of TE$_{01}$ mode which allows the propagation of cross-polarization signals is 212.6 GHz. Because 212.6 GHz is out of the design frequency range, this dimension simultaneously ensures a good insertion loss and a low intrinsic cross-polarization loss.
Then, each element of the OMT is optimized with Wasp-NET ~\cite{arndt2012wasp}separately. After a fully optimization for a best possible reflection performance, all elements of the OMT are combined. Then, some important dimensions are re-optimized with FE algorithm for an optimum performance of the whole component.
Finally, reflection losses lower than -25.0 dB for vertical polarization and -24.0 dB for horizontal polarization are realized in the OMT design.

The Band 4+5 OMT has also been fabricated by Kawashima Manufacturing in Japan~\cite{kawashima}. Because of the geometric design of this OMT, it can be fabricated by direct machining of aluminum in two split blocks, which relatively reduces the complexity of the fabrication. As shown in Fig.~\ref{fig:B4+5 OMT}(a), same as the Band 6+7 OMT, gold plating is applied on the surface of this OMT for a better transmission loss. 


As shown in Fig.~\ref{fig:B4+5 OMT}(c), this OMT has also been characterized by a Keysight PNA-X VNA at NICT Advanced ICT Device Laboratory. 
Because of the large fractional bandwidth, VDI WR-6.5 extenders are used for the measurement at 120--150 GHz, and OML WR-5.1 extenders are used for the measurement at 150 -- 210 GHz, respectively. In the measurement, two kinds of shim-type rectangular waveguide transitions (see Fig.~\ref{fig:B4+5 OMT}(b)) are used to connect the OMT with the VNA extenders. To reduce the extra reflection from the transitions, the dimensions of these shim-type transitions are fully optimized by Wasp-NET~\cite{arndt2012wasp}. As a result, the reflection losses for both WR-6.5 and WR-5.1 rectangular waveguide transitions are better than -35 dB. As shown in Fig.~\ref{fig:B4+5 OMT}(d), these transitions can be simply inserted as shims between the OMT and the VNA extenders. In comparison to traditional taper-type transitions, the insertion loss of shim-type transitions is negligible. Therefore, it is not necessary to compensate for the insertion loss of rectangular waveguide transitions in the measurement of this OMT, which highly reduces the complexity of the measurement. On the other hand, standard taper-type square waveguide transitions are used in this measurement, because it is hard to use only one waveguide step to achieve a good match between the square waveguide and the rectangular waveguide. Therefore, the insertion loss of the square waveguide transitions is also measured and compensated in the measurement result of the OMT.

The measurement result is presented in Fig.~\ref{fig:B4+5 OMT Meas}. Same as Band 6+7 OMT, this measurement is also based on the methodology presented in Ref.~\citenum{navarrini2021}. The measured transmission loss is around -0.4 -- -0.2 dB for horizontal polarization and -0.6 -- -0.4 dB for vertical polarization. These values are pretty close to the measurement result of ALMA Band 4 (125--163 GHz) OMT~\cite{asayama2009development}. Because of the narrower fractional bandwidth of ALMA Band 4 OMT, it uses a larger square waveguide input (WR-6.5, 1.65 $\times$ 1.65 mm), and the ohmic loss due to the finite conductivity is intrinsically lower than that of the Band 4+5 OMT reported in this proceeding. Therefore, the current result demonstrated that gold plating can improve the surface roughness of waveguide components for a better transmission loss.
As mentioned in the previous context, this measurement is also conducted at room temperature. The transmission loss is considered to be further improved when the OMT is cooled down, but it is hard to predict specific values due to the uncertainty of conductivity at cryogenic temperatures.
The measurement also shows that reflection loss is lower than -18.8 dB for H-pol and lower than -19.7 dB for V-pol in the whole frequency range. Although these values are several decibels worse than the design values in Fig.~\ref{fig:B4+5 OMT Meas}(c), the OMT still show an excellent reflection performance in this frequency range~\cite{asayama2009development}.
Furthermore, the cross-polarization loss of OMT is lower than -32.4 dB for both polarizations. This low level of cross-polarization demonstrated a good mechanical alignment between the two split blocks of this OMT.



\section{SIS mixers}
Low-noise SIS mixers which have wide RF bandwidths and wide IF bandwidths are the key devices for FINER receivers. NAOJ has been studying this technology for years, and several heterodyne receivers based on wideband SIS mixers have been demonstrated in recent years~\cite{kojima2017performance,kojima2018275,kojima2020wideband}. Soon, this technology will also be applied to ALMA Wideband Sensitivity Upgrade (WSU) to further extend the IF bandwidth of ALMA~\cite{carpenter2022alma2030}.
For this purpose, now NAOJ is developing a second version of ALMA Band 8 receiver with a wider IF bandwidth~\cite{kojima2024band8v2}. 
The key technology for low-noise wideband SIS mixers is high critical current density ($J_{c}$) SIS junctions which have low resistance to facilitate the extension of the RF and IF bandwidth of SIS mixers~\cite{kojima2017performance}.

\begin{figure} [h]
   \begin{center}
   \begin{tabular}{c} 
   \includegraphics[height=6cm]{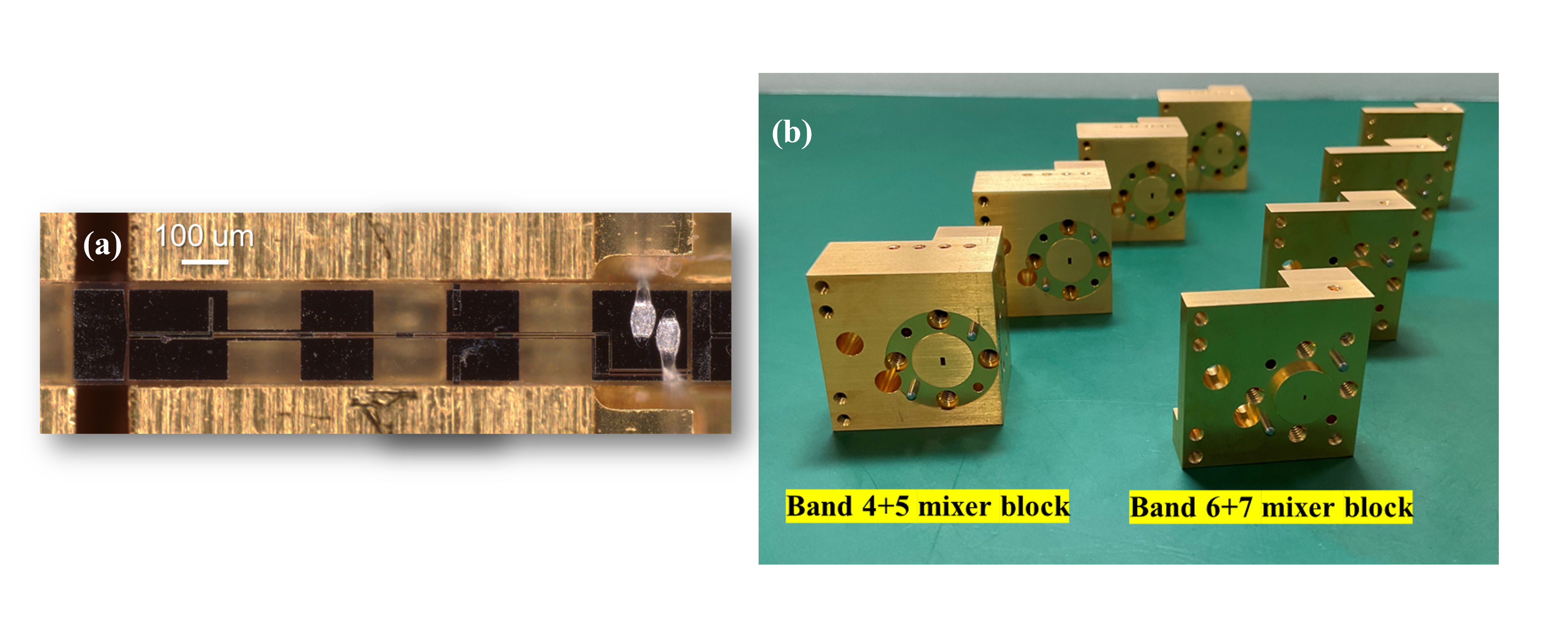}
   \end{tabular}
   \end{center}
   \caption[Mixer chip + blocks] 
   { \label{fig:Mixer chip and blocks} 
(a) The Band 6+7 SIS mixer chip. (b) The fabricated mixer blocks for Band 6+7 frequencies and Band 4+5 frequencies.}
   \end{figure}

In the design of the Band 6+7 SIS mixer, the RF circuit comprises a parallel-connected twin junction based on Nb/Al/AlNx/Al/Nb trilayers and RF matching circuits. In the meanwhile, to extend the IF bandwidth, the IF matching circuit is carefully designed with the consideration of both the output impedance of the SIS junction and the input impedance of the CLNA~\cite{kojima2017performance,kojima2018275}. As shown in Fig.~\ref{fig:Mixer chip and blocks}(a), a prototype Band 6+7 SIS mixer has been fabricated and tested with a drop-in CLNA module. The DSB receiver noise temperature of the prototype mixer is $\sim$ 30--35 K in the whole RF frequency band over 3--21 GHz IF frequencies, which is corresponding to 2–3 times of the quantum noise. In addition, as shown in Fig.~\ref{fig:Mixer chip and blocks}(b), the four mixer blocks which comprise of taper waveguide to match the SIS mixer with the WR-3.15 rectangular waveguide in waveguide circuits have also been fabricated. 
In the coming months, the full system evaluation for Band 6+7 receiver including wideband waveguide circuits, SIS mixers, LO system and backend digital spectrometer will be conducted.
On the other hand, the development of the Band 4+5 SIS mixer is still ongoing, it is expected to be completed by the first half of 2025.
\section{Conclusion}
This proceeding has summarized the development status of the key components for FINER heterodyne receivers at 210--360 GHz and 120--210 GHz. 

For the Band 6+7 receiver, the development of the wideband waveguide components almost has been completed. The measurement shows that the Band 6+7 corrugated horn has an excellent reflection performance, but the far-field beam patterns of this corrugated horn still need to be characterized. On the other hand, although the Band 6+7 OMT has a slightly high level of cross-polarization loss with a maximum of -24 dB, it still meets the scientific requirements of FINER project as polarization observations are out of the scope of the current scientific goals. The Band 6+7 2SB unit shows a slight degradation of amplitude imbalance in comparison to the design value, but because of the digital sideband separation function in the backend digital spectrometer for FINER, a high image rejection ratio is still expected in the final observation data. In addition, the prototype of wideband SIS mixer at 210--360 GHz has also been fabricated and tested, which shows a DSB receiver noise temperature around two to three times of the quantum noise, and the noise performance of this SIS mixer fully meets the requirement of FINER. Due to the steady development progress of this receiver, the full system evaluation for Band 6+7 frequencies will take place within the year of 2024. 

For the Band 4+5 receiver, the development of the OMT has been completed. The measurement shows an excellent reflection performance for both polarizations, and a cross-polarization loss far better than -30 dB. On the other hand, the corrugated horn and the 2SB unit at this frequency range are still under development, but they are expected to be completed within the year of 2024. Therefore, the remaining task is the development of the Band 4+5 SIS mixer. A large amount of effort is needed to achieve the required bandwidth at this frequency range. After the development of the Band 4+5 receiver, the development team of FINER is targeting to implement these two low-noise heterodyne receivers on LMT in late 2025 to early 2026.

\acknowledgments 
This work was supported by Japan Society for the Promotion of Science (JSPS) KEKENHI Grant No. 22H04939.
The experiment presented in this work was supported by NICT Advanced ICT Device Laboratory, and the authors would like to thank I. Watanabe at NICT for his kindly support of the experiment.

\bibliography{report} 
\bibliographystyle{spiebib} 

\end{document}